\documentclass[10pt,aps,prb,twocolumn,superscriptaddress,floatfix,showpacs,longbibliography]{revtex4-1}
\usepackage[utf8]{inputenc}
\usepackage{graphicx}
\usepackage{tabularx}
\usepackage[usenames,dvipsnames,table]{xcolor}
\usepackage[version=3]{mhchem}
\usepackage{braket}
\usepackage{upgreek}
\usepackage{hyperref}
\usepackage{amsmath, amssymb}
\usepackage[normalem]{ulem}
\usepackage{soul}
\usepackage{mathtools}

\definecolor{mark}{rgb}{0.85, 0.9, 1}
\definecolor{rred}{HTML}{CB4154}

\hypersetup{hidelinks}
\sethlcolor{mark}
\newcolumntype{Y}{>{\centering\arraybackslash}X}

\begin{document}

\title{Estimating Patterns of Classical and Quantum Skyrmion States}
\author{Vladimir V. Mazurenko}

\affiliation{Theoretical Physics and Applied Mathematics Department, Ural Federal University, Mira Str. 19, 620002 Ekaterinburg, Russia}

\author{Ilia A. Iakovlev}
\affiliation{Theoretical Physics and Applied Mathematics Department, Ural Federal University, Mira Str. 19, 620002 Ekaterinburg, Russia}

\author{Oleg M. Sotnikov}
\affiliation{Theoretical Physics and Applied Mathematics Department, Ural Federal University, Mira Str. 19, 620002 Ekaterinburg, Russia}

\author{Mikhail I. Katsnelson}
\affiliation{Radboud University, Institute for Molecules and Materials, Nijmegen, Netherlands\\
}
\date{\today}

\begin{abstract}
In this review we discuss the latest results concerning development of the machine learning algorithms for characterization of the magnetic skyrmions that are topologically-protected magnetic textures originated from the Dzyaloshinskii-Moriya interaction that competes Heisenberg isotropic exchange in ferromagnets. We show that for classical spin systems there is a whole pool of machine approaches allowing their accurate phase classification and quantitative description on the basis of few magnetization snapshots. In turn, investigation of the quantum skyrmions is a less explored issue, since there are fundamental limitations on the simulation of such wave functions with classical supercomputers. One needs to find the ways to imitate quantum skyrmions on near-term quantum computers. In this respect, we discuss implementation of the method for estimating structural complexity of classical objects for characterization of the quantum skyrmion state on the basis of limited number of bitstrings obtained from the projective measurements. 

\end{abstract}

\maketitle

\section{Introduction}


History of science unambiguously evidences that the development of new theoretical concepts and physical models is impossible without insights from experiments and observations, which is crucial for exploring physical systems of any scale, from planetary objects to atoms and less than atoms. One of the bright examples of this insight is magnetic measurements\cite{Smith} of a seemingly standard magnet, hematite $\alpha$-Fe$_2$O$_3$ performed by Smith in 1916, which, at the very end, paved the way to the concept of anisotropic inter-atomic interactions formulated by Igor Dzyaloshinskii \cite{Dzyaloshinskii} and Toru Moriya \cite{Moriya}. In his work Smith has found a ferromagnetic response by applying magnetic field perpendicular to the trigonal axis of $\alpha$-Fe$_2$O$_3$. Remarkably, this experimental observation playing the crucial role in the theory of the inter-spin interaction was done about 10 years before the concept of the electron spin itself was introduced. Further experiments \cite{Neel} confirmed Smith's findings for Fe$_2$O$_3$ and showed robust weak ferromagnetism in other antiferromagnets \cite{Borovik1, NiF2} (MnCO$_3$, CoCO$_3$, NiF$_2$) characterized absence of inversion symmetry center between nearest magnetic ions. This relation with crystallography assumes that we deal with an intrinsic property rather than something related to crystal lattice imperfections (impurities, violation of the stoichiometric composition and others) \cite{Vonsovskii}.  

By 1957, a critical mass of experimental data that evidence existence of a symmetry-dependent spontaneous magnetization in a number of antiferromagnets had been accumulated. In this year Igor Dzyaloshinskii proposed an elegant way to explain the magnetic moment by using the symmetry arguments based on the twisting of the spin arrangement due to inversion symmetry breaking \cite{Dzyaloshinskii}. For that anisotropic exchange interaction in the form ${\bf D}_{ij} [{\bf S}_i \times {\bf S}_j]$ was introduced into the spin Hamiltonian. Such an interaction is antisymmetric with respect to interchange of the spins and  favors noncollinear magnetic order. Toru Moriya has proposed a microscopic mechanism for this newborn coupling by developing its superexchange theory \cite{Moriya, Anderson1, Anderson2} with taking the spin-orbit coupling into account. Subsequently, the original Moriya's microscopic theory was consistently improved and refined in papers \cite{Aharony1, Rice, Aharony2}. Moreover, different numerical schemes based on the density functional theory calculations were developed to estimate Dzyaloshinskii-Moriya interaction (DMI) from first-principles calculations\cite{Solovyev, MI, PRB2005, KKML, Blugel}. All these methodological results facilitate modeling magnetic properties of completely different materials and make estimation of DMI to be a routine procedure in the modern computational physics. 

Remarkably, even after half century our understanding of the Dzyaloshinskii-Moriya interaction is far from being complete. It can be justified by the example that standard magnetic measurements techniques allow to estimate the magnitude and symmetry of DMI in concrete correlated materials, however, they do not provide the information on the DMI sign, which defines the local twist of the magnetic structure with respect to the atomic rearrangement due to inversion symmetry breaking. Such a problem was solved in 2014 when a new experimental
technique, suggested earlier in Ref. \onlinecite{Dmitrienko}, was developed in Ref. \onlinecite{DMIsign}.
This technique is based on the interference between two X-ray
scattering processes, where one acts as a reference
wave allowing to determine the sign of another. Experimentally, the sign of the DMI in correlated materials can be controlled with the occupation of the 3d shell. For instance, as it was shown in Ref.~\onlinecite{carbonates} there is a DMI sign change in the series of isostructural weak ferromagnets, MnCO$_3$ (DMI $<$ 0), FeBO$_{3}$ (DMI $<$ 0), CoCO$_3$ (DMI $>$ 0) and NiCO$_3$ (DMI $>$ 0). These experimental results agree with magnetic structure features obtained from the DFT calculations and can be explained Moriya's microscopic theory taking into account the occupation change of the correlated states.

Over time, it became clear that the scope of Dzyaloshinskii-Moriya interaction is not limited to antiferromagnetic insulators with weak ferromagnetism. 
It has been used for prediction of long-range spiral structures in certain magnets \cite{Dzyaloshinskii1964} without inversion symmetry. Later on such spiral structures were experimentally observed in metallic MnSi and FeGe magnets \cite{lebech,Ishikawa} and Fe$_{1-x}$Co$_{x}$Si alloys\cite{Beille1,Beille2} with the B20 crystal group. For these metallic systems, the period of the spiral structures varies in wide ranges from 175 \AA \, to 700 \AA. In turn, magnetic critical temperatures for stabilization of the spin-spiral ground state can be also very different and include technologically important regimes that are close to the room temperature. In addition, it was found that there is a complex interplay between magnetic properties and electronic structure of long-range spin-spiral metallic magnets. For instance, first-principles calculations \cite{FeSi} of the B20 crystal group systems have revealed a strong renormalization of the electronic spectra near the Fermi level due to the dynamical electron-electron correlations, which can also affect values of the magnetic moments as well as isotropic and anisotropic magnetic exchange interactions \cite{MnFeCo} and, therefore, should be taken into account when one describing these systems theoretically.    

The exploration of the materials hosting the DMI spin spirals played an important role in establishing new research field of topologically-protected magnetic structures. It was first shown theoretically \cite{Bogdanov} in 1989 and then confirmed experimentally \cite{skyrm_exp1,skyrm_exp2} that the DMI is responsible for forming long-range topologically protected chiral structures, magnetic skyrmions in metallic ferromagnets. The possibility to stabilize and manipulate skyrmions with magnetic and electric fields at the room temperature makes them very promising in numerous technological applications including next-generation memory devices and quantum computing \cite{sk_qubit}. Undoubtedly, further strides in the field of the magnetic skyrmions as well as creating new technologies that will make of use topological properties of the materials require implementation of the most advanced techniques for generation, detection, exploration and control. In this respect machine learning and quantum computing are of special interest. While the former allows to automatically classify and characterize magnetic structures the latter facilitates imitation of new phases of matter including topological ones. 

Keeping in mind the recent progress in developing machine learning and quantum computing techniques for scientific research in this review paper we first focus on the latest activity concerning the implementation of computing methods for exploration of the magnetic skyrmion phases. A special accent will be given on the recently introduced renormalization procedure for calculating structural complexity of an object \cite{PNASComplexity}, which allows straightforward estimation of the phase boundaries in non-collinear magnets in a purely unsupervised manner by using a few magnetic snapshots of the system in question. The second part of the paper is devoted to theoretical analysis of the quantum skyrmions that are ground states of quantum systems with Dzyaloshinskii-Moriya interaction. Our abilities in simulation of such quantum states with classical computers are limited due to the exponential growth of the Hilbert space with the number of particles. At the same time quantum computing can be considered as the most promising technology for further exploration of the quantum skyrmion states. In this respect the development of approaches for certification and identification of quantum states of large-scale quantum systems are in demand and attract considerable attention. We discuss the generalization of the procedure for calculating structural complexity of object onto the case of quantum states and report on the classification of the quantum phases in DMI magnet on the basis of the bitstrings obtained after a limited number of projective measurements. The problems concerning imitation of the quantum skyrmions  with quantum computers are discussed. 

\section{Classical skyrmions}

\begin{figure}[!b]
	\includegraphics[width=\columnwidth]{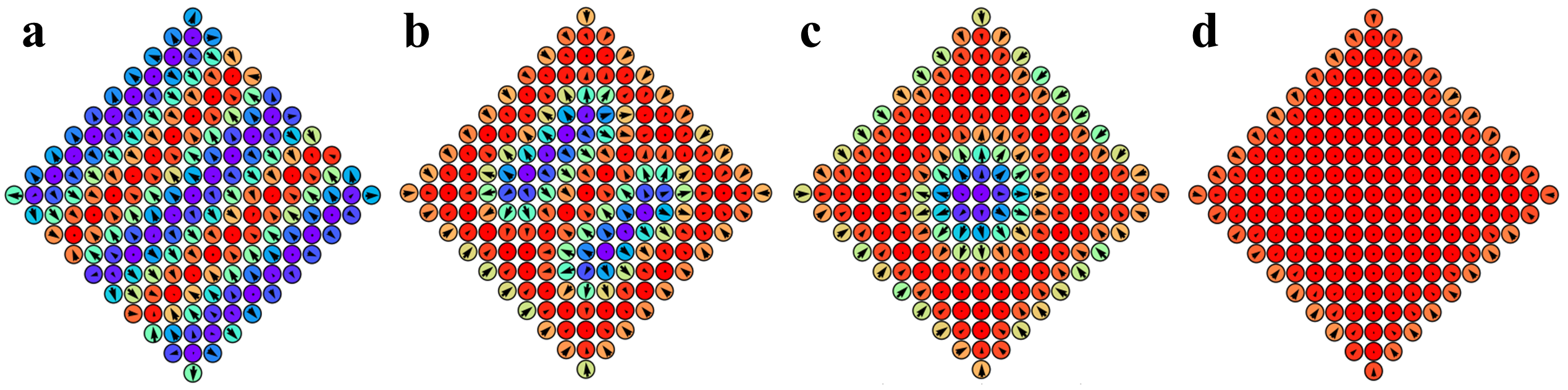}
	\caption{\label{plaquette} (Color online) Monte Carlo solutions of the spin model, Eq.~\eqref{Ham} on the rhombic plaquette, corresponding to the spin spiral (a), bimeron (b), skyrmion (c) and ferromagnetic states. This figure is reproduced from Ref. \onlinecite{bimeron}. (c) [2018] American Physical Society.}
\end{figure}

A typical spin Hamiltonian used to simulate skyrmion structures can be written in the following form: 
\begin{eqnarray}\label{Ham}
H = \sum_{ij} J_{ij} {\bf S}_i {\bf S}_j + \sum_{ij} {\bf D}_{ij} [{\bf S}_i \times {\bf S}_j] + B \sum_i S_{i}^z
\end{eqnarray}
where $J_{ij}$ and ${\bf D}_{ij}$ are the isotropic interaction and Dzyaloshinskii-Moriya vector, respectively, ${\bf S}_{i}$ is a unit vector along the direction of the $i$th spin and $B$ denotes the out-of-plane magnetic field.   To stabilize a skyrmion state for Hamiltonian (\ref{Ham}) the Dzyaloshinskii-Moriya interaction for each bond should be in-plane. In our work we consider ${\bf D}_{ij}$ that points in the direction perpendicular to the bond between neighboring $i$ and $j$ sites. At low temperatures, different phases realized with Hamiltonian Eq.\ref{Ham} can be identified using conventional techniques, that is, calculating skyrmion number and spin structure factors.  The skyrmion number (topological charge) $Q$ is defined as
\begin{equation}\label{Q}
Q=\frac{1}{8\pi}\sum_{\langle ijk\rangle}\bf{S}_i\cdot[\bf{S}_j\times\bf{S}_k],
\end{equation}
where the summation runs over all nonequivalent elementary triangles that connect neighboring
$i$, $j$, and $k$ sites. In the case of skyrmions of a few atoms in size~\cite{pureDMI} one can use an approach proposed in Refs.~\onlinecite{BergLuscher} and \onlinecite{HeoBlugel}. In turn, the spin structure factors are given by
\begin{equation}\label{hi1}
\chi^\parallel_{\textbf{q}}=\left\langle {S}^z_{\textbf{q}} {S}^z_{\textbf{-q}}\right\rangle,
\end{equation}
\begin{equation}\label{hi2}
\chi^\perp_{\textbf{q}}=\left\langle {S}^x_{\textbf{q}} {S}^x_{-\textbf{q}}\right\rangle + \left\langle {S}^y_{\textbf{q}} {S}^y_{-\textbf{q}}\right\rangle,
\end{equation}
where $\textbf{q}$ is the reciprocal space vector.

\begin{figure}[!t]
\begin{center}
\includegraphics[width=\columnwidth]{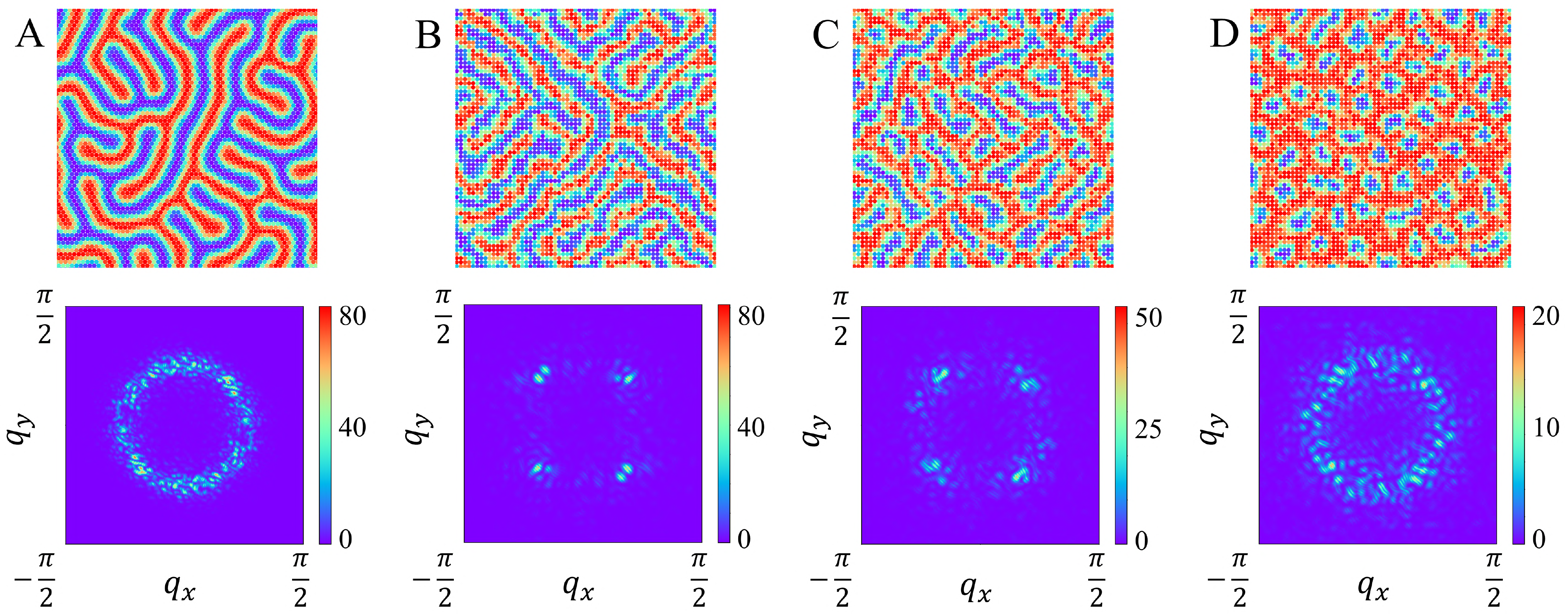}
\end{center}
\caption{(Color online) (A) Magnetic labyrinth on a triangular lattice at low temperature ($T=0.02J$), (B) spin spirals, (C) mixed skyrmion-bimeron magnetic configuration and (D) pure skyrmions on a square lattice at high temperature ($T=0.4J$), and the corresponding spin structure factors. This figure is reproduced from Ref. \onlinecite{PNASComplexity}. (c) [2020] National Academy of Sciences.}
\label{struct}
\end{figure}

As it was previously shown~\cite{skyrm_exp2, bimeron}, phase diagram of such system (Eq.\ref{Ham}) consists of three clear phases: spin spirals, skyrmion crystal and ferromagnetic state, and two significant intermediate regions, namely, skyrmion-bimeron state and skyrmion gas. Some examples of these spin textures are presented in Fig.~\ref{plaquette}. By bimeron here we mean a spin texture composed of two half-disk meron domains having $Q=\pm1/2$ divided by neutral rectangular stripe domain~\cite{Ezawa}. Such quasi-particle can be associated with either elongated skyrmions or broken helix segments, since its length strongly depend on the Hamiltonian parameters. We should note that term bimeron is rather used to describe particles, composed of a pair of merons of different vorticity~\cite{varb1, varb2, varb3}. However, to observe them one has to change the symmetry of the DM vector. The detailed description of these quasi-particles is given in Ref.~\onlinecite{bimreview}.

Unfortunately, the skyrmion number and spin structural factors are very sensitive to temperature and give us inappropriate results even in case when the spin structures still remain visually recognisable~\cite{our2Dsk}(see Fig.~\ref{struct}). This fact aroused significant interest in the development of machine methods for conducting phase classification in this system. Below we will discuss such techniques.  

\subsection{Complexity}

\begin{figure}[!t]
	\includegraphics[width=\columnwidth]{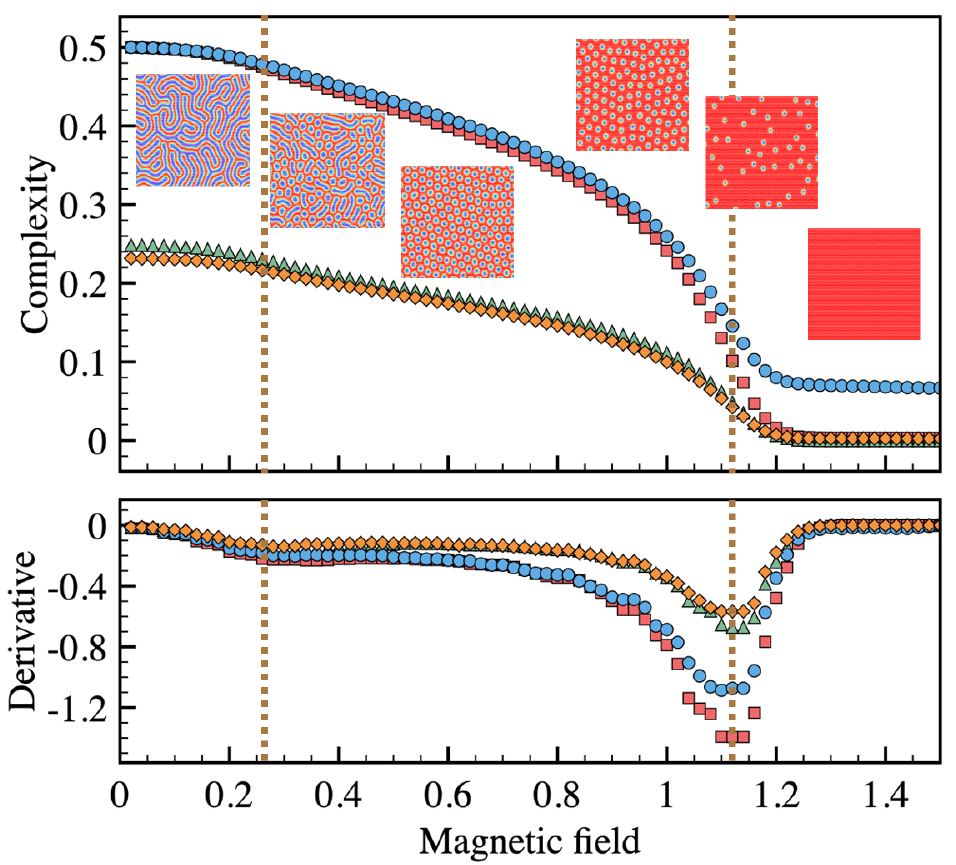}
	\caption{\label{sk_comp} (Color online) (Top) Structural complexity calculated for the classical two-dimensional triangular lattice magnetic configurations obtained with Hamiltonian Eq.\eqref{Ham} as a function of external magnetic field. The error bars are smaller than the symbol size. (Bottom) Complexity derivative we used for accurate detection of the phases boundaries. Squares and circles correspond to the low ($T=0.02J$) and high ($T=0.4J$) temperature configurations obtained with using all spin components, meanwhile triangles and diamonds represent results obtained with using only $z$ component. All the results are obtained for $|D|=J$, only the interaction between the nearest neighbours is taken into account.}
\end{figure}

One of the basic concept which is usually used to analyze various patterns, systems and processes is their structural complexity. Although being intuitively clear since it reflects human's perception of reality, this value is very difficult to describe quantitatively. However, a lot of domains from geology to social sciences require a robust mathematical notion that properly reflects complexity of hierarchical non-random structures. Despite numerous attempts to give a formal definition of this quantity \cite{badii,MMC,GL,lloyd,CarrNetwork,Giulio,DeGiuli,beauty}, our understanding of these matters is still far from being complete. Recently, some of us have proposed an easy to compute, robust and universal definition of structural (effective) complexity based on inter-scale dissimilarity of patterns~\cite{PNASComplexity}. Besides meeting an intuitive perception of what is ``complex'' and what is ``simple'', this measure has been shown to be a suitable tool for determining phase transitions in various types of systems, including skyrmion structures.   

In its simplest form, the algorithm to compute structural complexity of a given magnetic configuration consisting of $L\times L$ atoms can be formulated in the following way \cite{PNASComplexity}: at each iteration, the whole system is divided into blocks of $\Lambda \times \Lambda$ size, and each block is substituted with a single spin which is calculated as ${\bf s}_{ij} (k) = \frac{1}{\Lambda^2} \sum_{l} \sum_{m} {\bf s}_{\Lambda i+m, \Lambda j+l} (k-1)$, where the $lm$ indices enumerate the spins belonging to the same block, and $k$ is the number of iteration. Then, one can compute overlaps between patterns separated by one step of such an averaging procedure:
\begin{gather}
O_{k, k-1} =\frac{1}{L^2}\sum_{i=1}^{L} \sum_{j=1}^{L}  {\bf s}_{ij} (k)\cdot {\bf s}_{ij} (k-1),
\end{gather}
with $k=0$ corresponding to the original pattern, and $O_{k,k}$ is an overlap of the pattern at scale $k$ with its own self. Defining structural complexity $\cal C$ as an integral characteristic accounting for features emerging at every new scale, we obtain
\begin{gather}
\label{eq:Complexity}
{\cal C} = \sum\limits_{k=0}^{N-1} {\cal C}_k =  \sum\limits_{k=0}^{N-1} |O_{k+1,k} - \frac12\left(O_{k, k}+O_{k+1, k+1}\right)|,
\end{gather}
where $N$ is the total number of averaging steps.

Fig.\ref{sk_comp} shows an example of the implementation of the structural complexity approach to the skyrmion problem. More specifically, it gives the resulting dependence of structural complexity on magnetic field for two-dimensional triangular lattice system described by Hamiltonian Eq.\eqref{Ham}. Remarkably, for each value of $B$ the complexity appears to be very robust, fluctuating within $0.01\%$ error range for independent Monte Carlo runs. It means that one can safely use a single magnetization image for each magnetic field value to define the complexity. The extrema of complexity derivatives $d{\cal C}/dB$ reflect very well both the melting of spin spirals (magnetic labyrinths) into skyrmion crystals, with the transition point being exactly the bimeron phase, as well as the transition between skyrmion crystals and ferromagnets. 

Recently, the structural complexity was used to find the phase boundary between self-induced spin-glass and noncollinear magnetically ordered states in elemental Nd at low temperatures \cite{Nd}.

\begin{figure}[!b]
	\includegraphics[width=0.9\columnwidth]{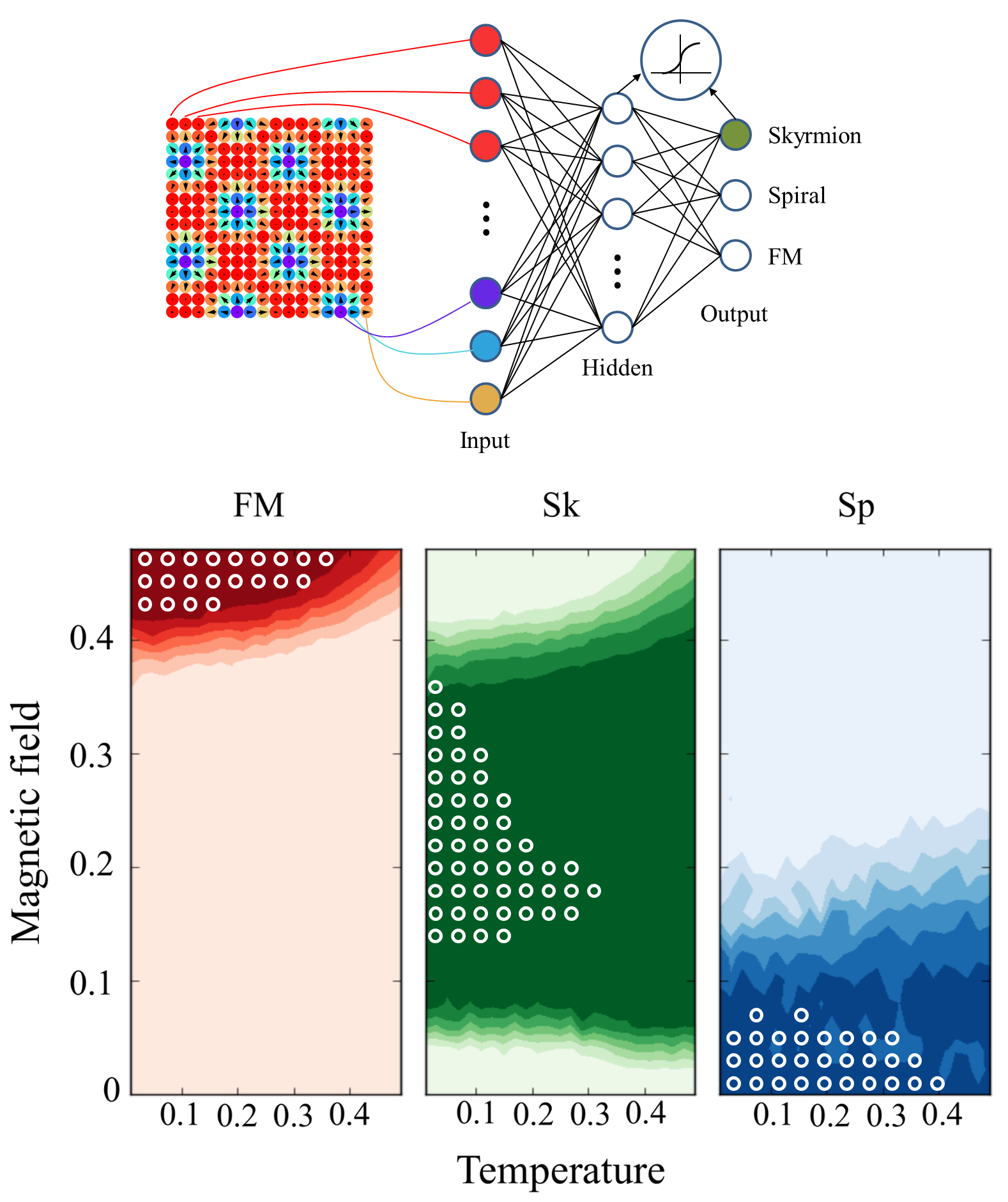}
	\caption{\label{2d_nn} (Color online) (Top) Schematic representation of the machine learning process. Neural network with single hidden layer of sigmoid neurons performs phase classification based on $z$ components of spins of the magnetic configuration. (Bottom) Phase triptych obtained by using the neural network with 64 hidden neurons for $|D|=0.72J$. White circles denote the phase boundaries defined with the spin structure factors. This figure is reproduced from Ref. \onlinecite{our2Dsk}. (c) [2018] American Physical Society.}
\end{figure}

\subsection{Machine learning methods for phase classification}
By the construction the method for calculating structural complexity can be classified as unsupervised one, since it does not use any apriori information on the system in question. At the same time, we would like to stress the advances of various supervised techniques that involve a learning with pre-prepared and labeled data. After the inspiring work of Carrasquilla and Melko~\cite{Melko}, who had demonstrated the ability of neural networks to define phase transitions in magnetic systems, significant efforts have been made in this field. Here we give a brief overview of such approaches aimed to study the properties of the magnetic skyrmions. 

\begin{figure}[!b]
	\includegraphics[width=0.9\columnwidth]{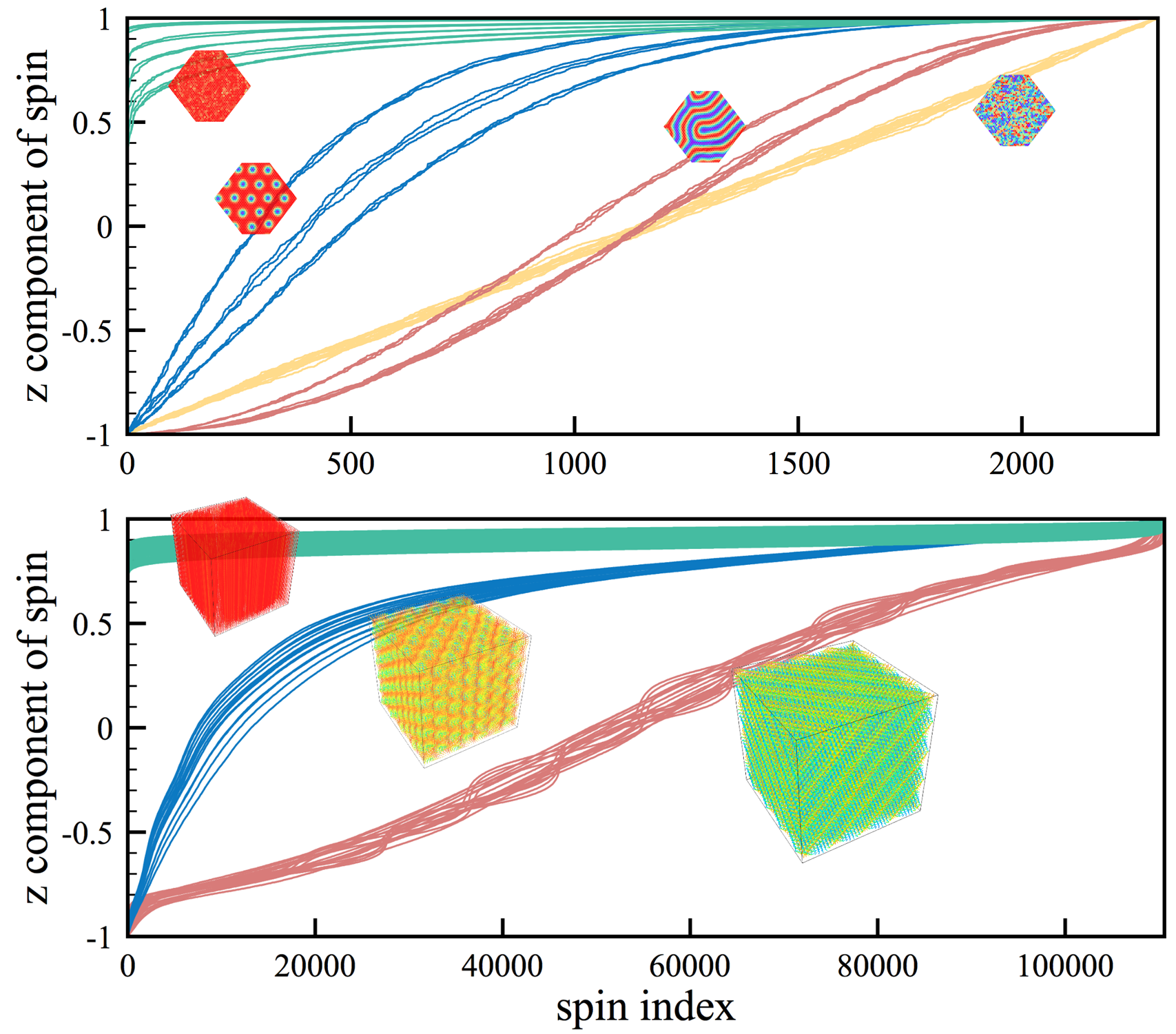}
	\caption{\label{fig_profiles} (Color online) Magnetization profiles of configurations belonging to different phases obtained with the spin Hamiltonian Eq.~\eqref{Ham} on the two-dimensional triangular lattice (top) and three-dimensional cubic lattice (bottom). This figure is reproduced from Ref. \onlinecite{our3Dsk}. (c) [2019] American Physical Society.}
\end{figure}

In Ref.~\onlinecite{our2Dsk} we have demonstrated the possibility to use machine learning algorithms for exploration phases of  non-collinear magnets that can host skyrmionic structures (see Fig.~\ref{2d_nn}). In the work\cite{our2Dsk}, a square lattice system described by a spin Hamiltonian \eqref{Ham} was considered. Given the fact, that all presented spin textures have distinct magnetization profiles, we decided to use only $z$ components of spins as an input for machine learning algorithms. We have found out, that a simple single-hidden-layer feed-forward neural network (FFN) with only 64 hidden neurons, trained on a moderate number of clear-phase Monte Carlo configurations, was able to successfully reproduce the entire phase diagram, including intermediate regions. Moreover, it demonstrated good results on unseen data, namely, high-temperature configurations, larger skyrmions and configurations obtained for triangular lattice system. Unfortunately, we found that such a network relies mostly on total magnetization and therefore cannot distinguish spin spirals and paramagnetic state. However, such an inconvenience can be easily overcome by simple sorting of the input vector even in case of 3D systems~\cite{our3Dsk} (see Fig.~\ref{fig_profiles}). It was also shown, that standard machine learning techniques like $k$-nearest neighbours, nearest centroids and support-vector machine work well in case of all clear phases and paramagnetic state.

\begin{figure}[!b]
	\includegraphics[width=0.9\columnwidth]{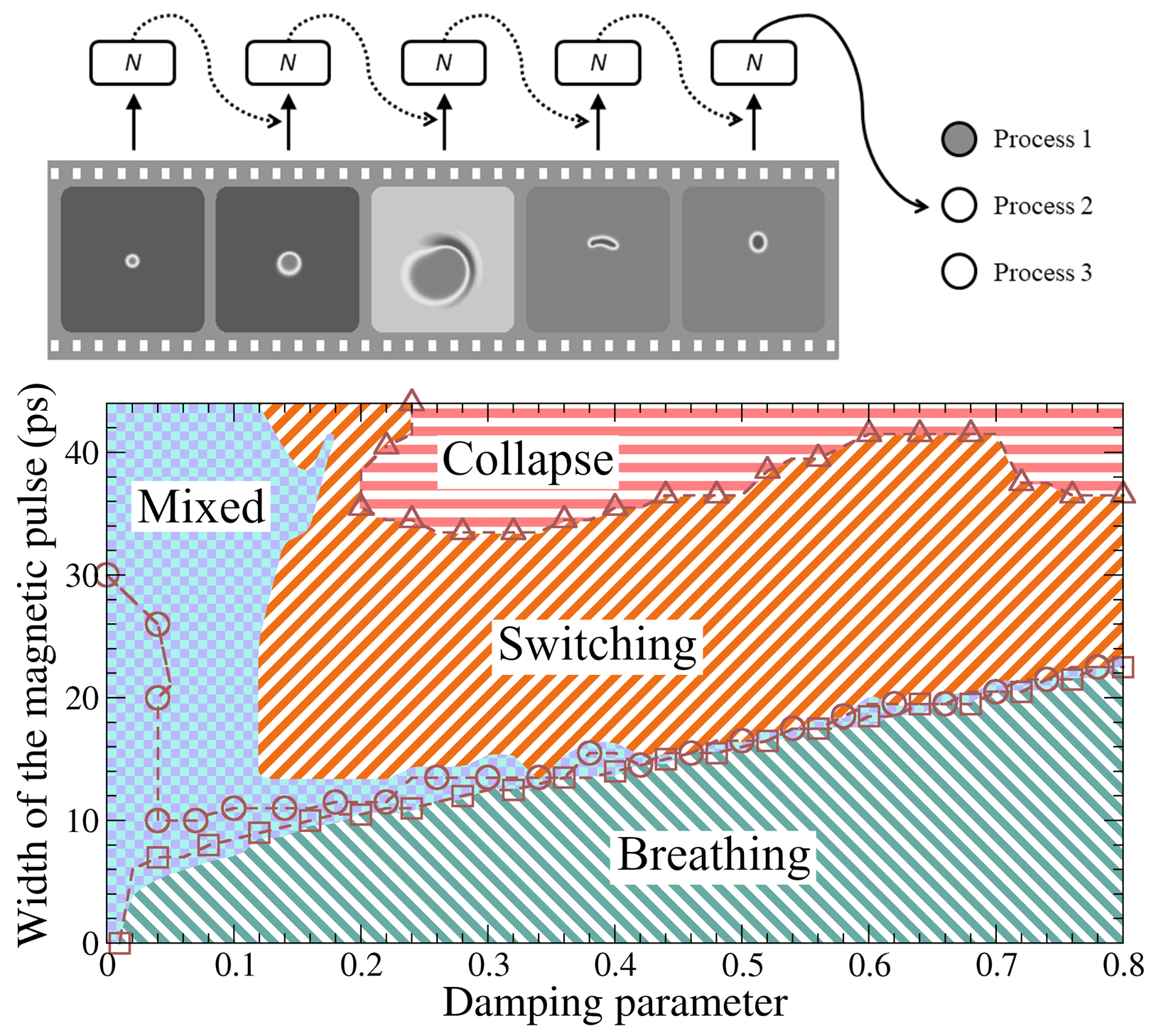}
	\caption{\label{fig_rnn} (Color online) (Top) Illustration of idea of the ultrafast skyrmionic process recognition. Magnetization dynamics is used frame by frame as an input for recurrent neural network providing the process classification. (Bottom) Obtained process diagram with the $\theta = 40^\circ$ magnetic pulses. Phase boundaries determined by means of RNN are indicated by brown dashed lines. This figure is reproduced from Ref. \onlinecite{ourRNN}. (c) [2019] American Physical Society.}
\end{figure}

Later, more complicated convolutional neural networks (CNN) were used to study the effect of uniaxial magnetocrystalline anisotropy pointing in the $z$ direction on the phase diagram of a disk-shaped system~\cite{CNNdisk}, and to construct detailed phase diagrams for skyrmion systems including intermediate regions and paramagnetic state~\cite{CNNRosales}. Moreover, it was shown that such an architecture is able to not only determine phase boundaries but also restore various parameters. The authors of Ref.~\onlinecite{Han} have demonstrated, that a CNN trained on ground state configurations successfully recovers the chirality and magnetization of a given spin texture, as well as the temperature and external magnetic field at which it was stabilized. It is interesting to note that the accuracy of the algorithm remains remarkably high in presence of disorder caused by the randomly generated site-dependent uniaxial anisotropy. The authors of Ref.~\onlinecite{MatthiesQ} have addressed an important problem of finding a topological charge of a given system based on its time-integrated space-dependent magnetization. They demonstrated that, being analytically inaccessible, this quantity can be extracted using a CNN with almost 100\% accuracy. It was shown, that such an approach works well on systems of different confined geometries, including random islands, which looks very promising from the point of view of potential application to real experimental data.

Recently, considerable attention has also been paid to the dynamic properties of skyrmion structures. Some of us have shown that the simplest recurrent neural network (RNN) is able to automatically detect different processes occurring with an isolated skyrmion under the influence of picosecond magnetic field pulses~\cite{ourRNN} (see Fig.~\ref{fig_rnn}). Such an approach is promising as a technique which performs an autonomous control of the system's dynamics in case of a prototype of skyrmionic data storage elements~\cite{HeoBlugel}. The authors of Ref.~\onlinecite{WangRNN} have studied a dynamic phase diagram of a particle model for skyrmions in metallic chiral magnets with using CNN-RNN architecture. It was shown, that the network is able to not only draw the correct phase boundaries but also define the exact number of the order parameters of the system in question.

\section{Quantum skyrmions}

The progress in the development of experimental techniques \cite{skyrm_exp1,skyrm_exp2} for the observation of magnetic skyrmions, topologically protected spin structures, poses new challenges for the theory and numerical simulations of ordered magnetic phases \cite{Back}. Nowadays, skyrmions are mostly discussed in the context of spintronics, where these stable magnetic structures are proposed as bits in magnetic memory devices \cite{Fert}. The need to store more and more information requires the development of ultra-dense memories. This fact motivates the investigation of skyrmions of a nanoscale size, with recent significant progress. Skyrmions with the characteristic size of a few nanometers have already been observed in real experiments \cite{Heinze} and were theoretically predicted in magnets with DMI\cite{Ivanov}, in frustrated magnets \cite{Leonov}, as well as in narrow band Mott insulators under high-frequency light irradiation \cite{Stepanov}, and others. On such small characteristic length scales compared to the lattice constant, quantum effects cannot be neglected. Given this, the numerical study of classical spin models can no longer be considered as an comprehensive solution of the problem. Quantum fluctuations play a crucial role, because, strictly speaking, the spin itself is a quantum characteristic of an electron. 

A common way to approach this problem is to force the quantum system to behave as a classical one. As the result, description of a quantum skyrmionic problem is either done semiclassically assuming that the magnetization dynamics is dominated by classical magnetic excitations that emerge on top of the symmetry-broken ground state of the system \cite{Balents}, or by means of the Holstein-Primakoff transformation, which only allows to compute quantum corrections to the classical solution \cite{Fernandez}. Besides, in paper \cite{Rosch} topological states of small clusters embedded in the ferromagnetic environment were investigated.  

\begin{figure}[!t]
	\includegraphics[width=\columnwidth]{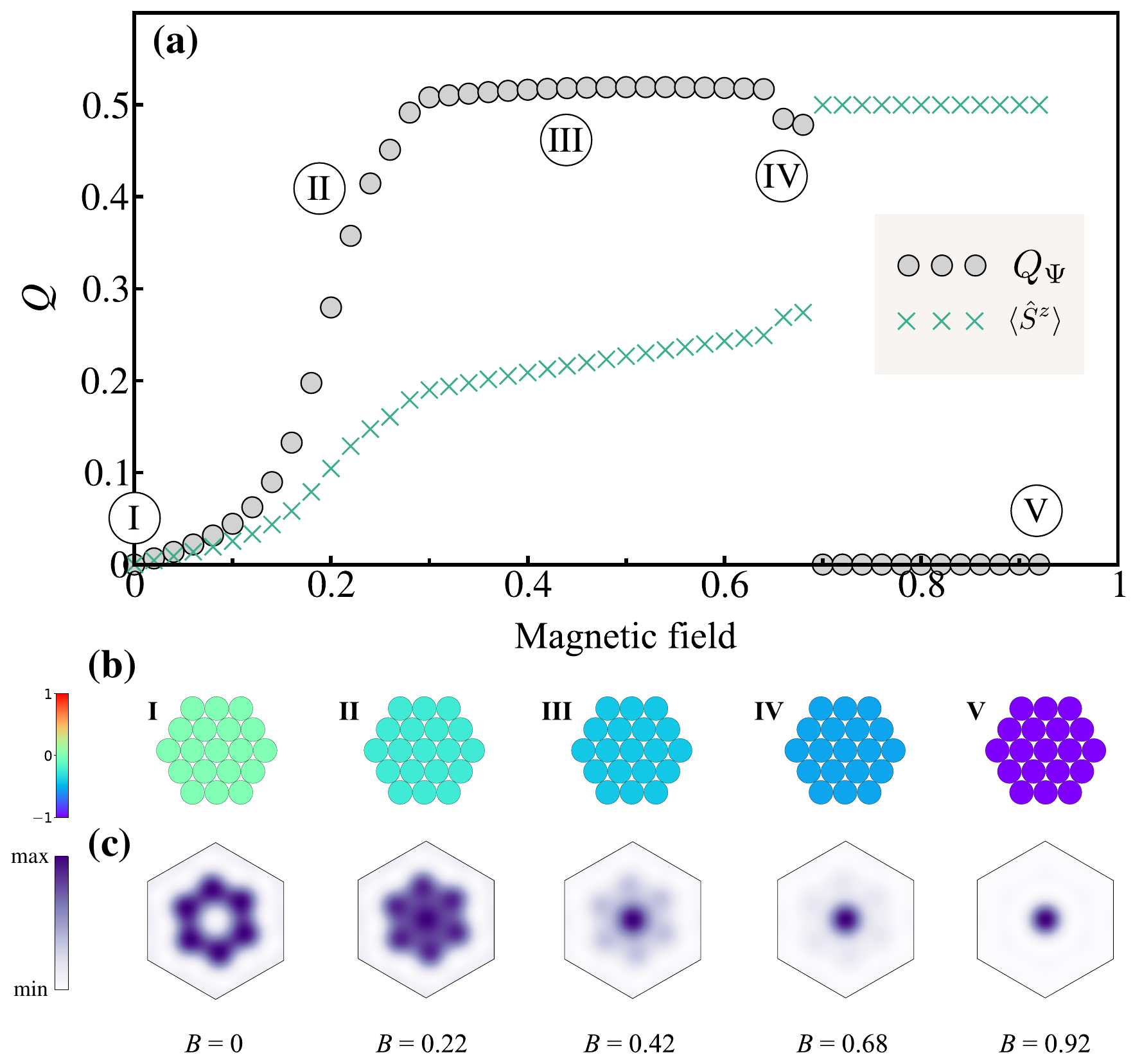}
	\caption{\label{qsk}(Color online) (a) Calculated scalar chirality and magnetization as functions of the magnetic field. (b) Magnetization density. (c) Calculated spin structural factors. This figure is reproduced from Ref. \onlinecite{Quantum_skyrmion}. (c) [2021] American Physical Society.}
\end{figure}

Recently, some of us have developed an approach for a  characterisation of the quantum skyrmion state \cite{Quantum_skyrmion} in an infinite magnetic systems for which, in contrast to the classical case, the magnetization density is uniform. For that, the following spin model defined on the 19-spin supercell with periodic boundary conditions was considered,
\begin{eqnarray}
\hat H = \sum_{ij} J_{ij} \hat {\bf S}_i \hat {\bf S}_j + \sum_{ij} {\bf D}_{ij} [\hat {\bf S}_i \times \hat {\bf S}_j] + B \sum_i \hat S_{i}^z.
\label{qHam}
\end{eqnarray}
Here $\hat {\bf S}_i$ is the spin-1/2 operator.
For characterization of the quantum ground states obtained at different magnetic field values, the local three-spin correlation function, $Q_{\Psi} = \frac{N}{\pi} \langle \hat {\bf S}_{1} \cdot [\hat {\bf S}_{2} \times \hat {\bf S}_{3}]  \rangle$ defined on neighboring lattice sites (here $N$ is the number of non-overlapping triangles in the supercell) was used. Such a correlator gives information about the topology of the entire quantum system, for instance, from Fig.\ref{qsk} one can see that the scalar chirality is characterized by non-zero constant value for the magnetic fields $0.3 < B < 0.66$, which clear signature of the quantum skyrmion phase.   

 Theoretically, exponential growth of the Hilbert space is the main factor preventing one from simulating larger systems than 19-site clusters discussed in Ref.~\onlinecite{Quantum_skyrmion} and from exploring quantum skyrmions of different kinds and sizes. The topological spin structures such as skyrmions emerge as the result of a competition between different magnetic interactions, leading to a magnetic frustration that restricts the applicability of quantum Monte Carlo methods due to the notorious sign problem \cite{QMCsign}.  In turn, exact diagonalization (ED) based methods have a severe restriction on the cluster size. For instance, for spin-1/2 Heisenberg-type models, the current limit is 50 lattice sites \cite{Lauchli}. Account of the anisotropic terms such as Dzyaloshinskii-Moriya interaction leads to mixing of the sectors of the Hamiltonian with different total spins, which significantly limits our opportunities to use symmetry of the system in question to reduce the size of the Hamiltonian matrix. Thus, this supercell of 19 sites with isotropic and anisotropic exchange interactions between spins defines the current limit on the system size for simulating quantum skyrmions with exact diagonalization approach. 

In this tough situation quantum computing provides a promising alternative to the standard approaches aimed at the search for ground and low-lying excited states of quantum Hamiltonian. Over the last decades there has been a fantastic progress in quantum computing on the level of constructing complete operating devices of up to 65 qubits with on-line access as well as in developing numerous algorithms for different fields of research including material science and condensed matter physics \cite{Kandala}. The first attempt \cite{Arute} of the Google team to demonstrate a quantum supremacy by the example of quantum chaos states have additionally heat up the interest of scientific community to the work on quantum states that are significantly delocalized in Hilbert space. Our preliminary results show that the quantum skyrmion and spin spiral states being solutions of the quantum spin model, Eq.\ref{qHam} are significantly spread over Hilbert space, which is an indication that their further theoretical exploration may be effective with quantum computers. However, it calls for development of the methods for characterization and identification of the quantum states. Below we will discuss such a procedure based on the estimating structural complexity of the bit-string patterns.

\subsection*{Calculating inter-scale dissimilarity of bit-string arrays}

\begin{figure}[!t]
	\includegraphics[width=0.5\columnwidth]{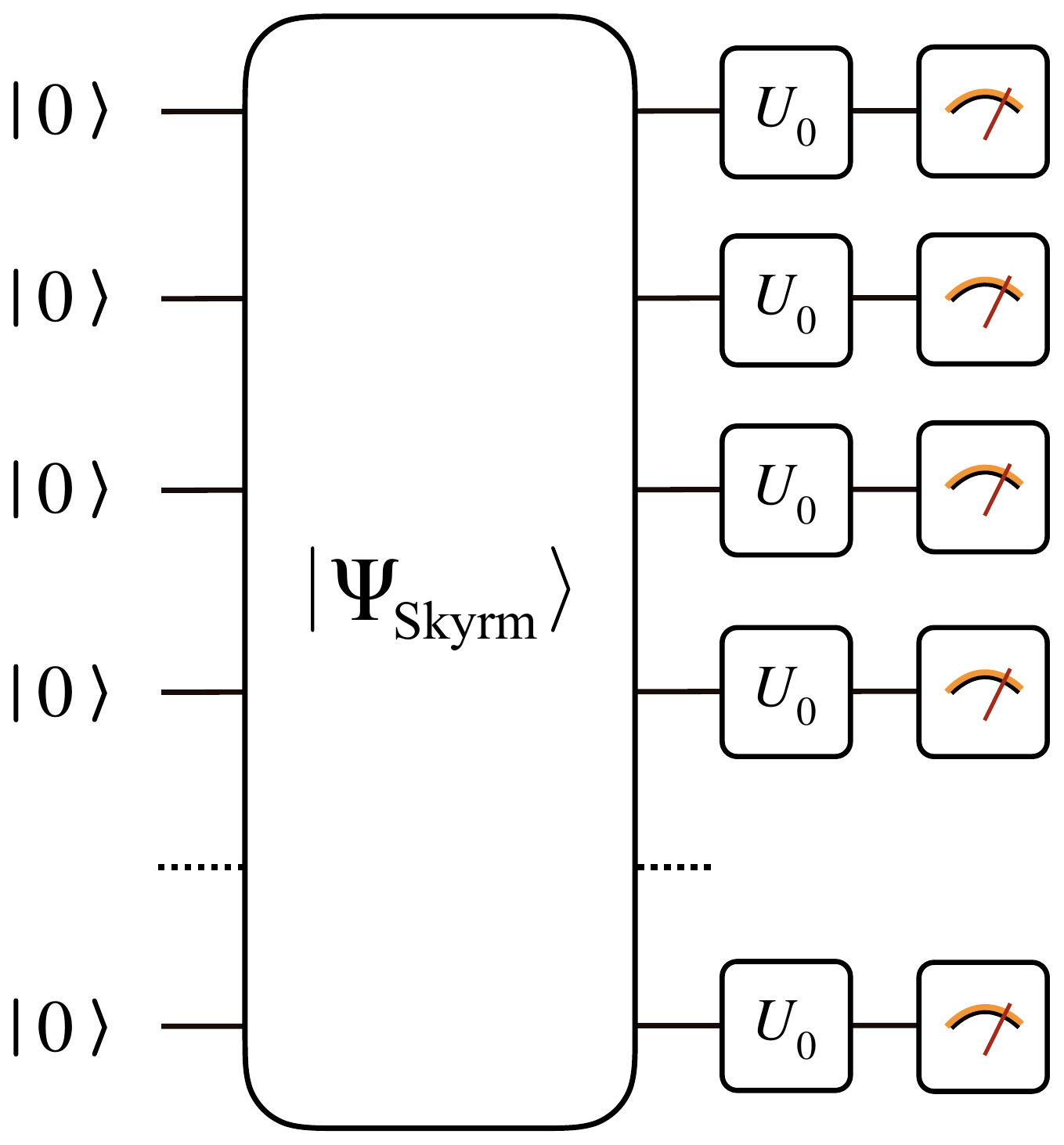}
	\caption{\label{randombasis} (Color online) Schematic representation of initialization of the quantum skyrmion state on a quantum simulator(computer). The measurement basis is chosen with one-qubit rotational gate U$_0$ that is described in the text.}
\end{figure}

Generally, the exploration of the magnetic skyrmions with quantum devices or simulators should include the following steps. First of all, one needs to define a quantum circuit that transforms the initial trivial quantum state $|000..0\rangle$ into desire $\Psi_{\rm Skyrm}$ (Fig.\ref{randombasis}) being the ground state of the quantum Hamiltonian, Eq.\ref{qHam}. There are two alternative solutions of this problem. (i) One can use a variational approach \cite{Troyer,Kandala,RL} in which the quantum state is represented with a fixed sequence of one- and two-qubit gates. The parameters of these gates are tuned to get the best approximation of the desire state. Unfortunately, there is no a universal sequence of gates that can approximate the ground state of an arbitrary Hamiltonian. Another problem is that one obtains not a true target state but only its approximation, which is of crucial importance for certain tasks. (ii) On the other hand, in the case of the small-size problems, the exact decomposition of the quantum ground state over basis functions can be found with exact diagonalization. It allows to employ Least Significant Bit (LSB) procedure~\cite{LSB} or similar procedure to find a sequence of gates realizing such a state. In the case of the 19-site quantum skyrmion ground state the LSB circuit contains thousands of gates, which is appropriate for creation and manipulation of such a state on a quantum simulator that imitates quantum logical operations on classical computer, but not on a real quantum device subjected to decoherence. For the latter one needs to develop new approaches aiming at generating quantum circuits that are as compact as possible. In this work we explore the quantum skyrmion state on the quantum simulator by using the LSB procedure. 

When quantum is initialized with target state, the measurements in a basis are performed (Fig.\ref{randombasis}). As it was shown in our previous work~\cite{Dissimilarity} to characterize a quantum state with dissimilarity procedure it is necessary to perform measurements in at least two bases because measurements in one fixed basis give the information only about amplitudes of wave function coefficients, but not the local phases. Following Ref.~\onlinecite{Dissimilarity} the measurements in $\sigma^z$ and random bases were performed. For each basis the measurement outputs are concatenated together into one string which can be considered as binary array of length $L = N\times{}N_{shots}$.
\begin{equation*}
   {\large \underbracket[1pt]{01011010}_{\text{\normalsize shot 1}}\underbracket[1pt]{00111010}_{\text{\normalsize shot 2}}\underbracket[1pt]{01100100}_{\text{\normalsize shot 3}}\underbracket[1pt]{01111010}_{\text{\normalsize shot 4}}}
\end{equation*}

After sampling of bit-string arrays, we estimate its patterns using the procedure presented in Ref.~\onlinecite{Dissimilarity}. Below we reproduce the main steps of this procedure.

\begin{figure}
    \includegraphics[width=0.9\columnwidth]{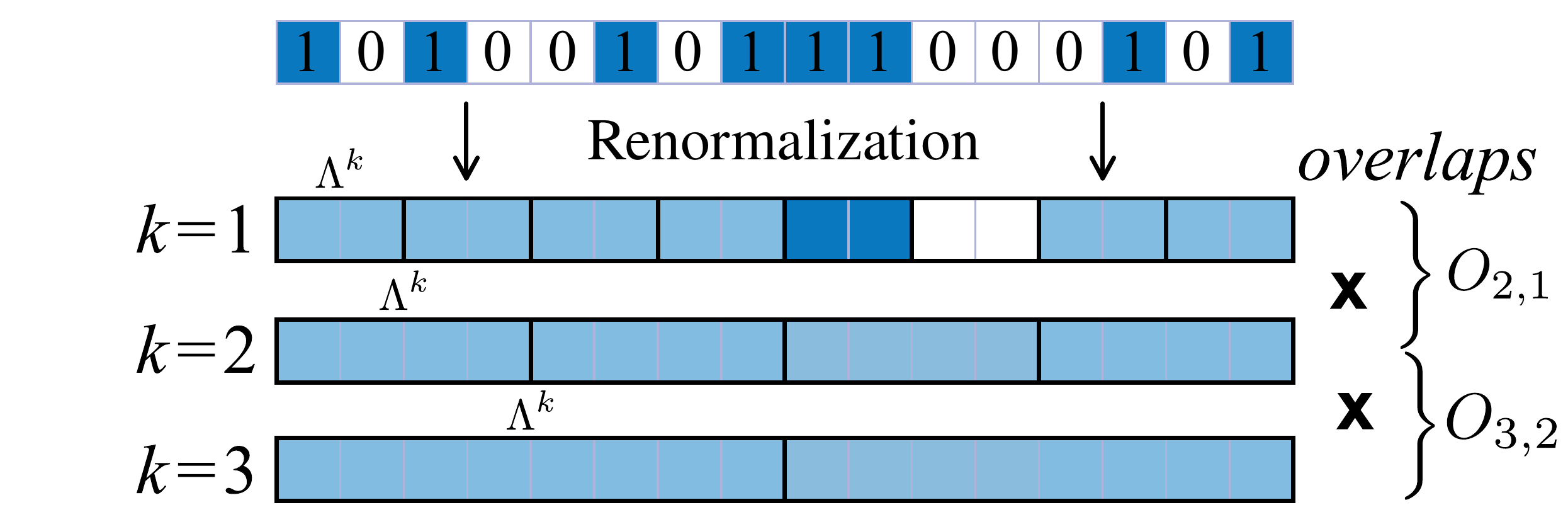}
    \caption{(Color online) Schematic representation of the renormalization procedure for bitstring array. First three steps are shown. At each step the elements belonging to the same block (black rectangle) are averaged. \label{renorm}}
\end{figure}

 At every step of coarse-graining $k$, a vector of the same length $L$ is constructed as
\begin{eqnarray}
b_{i}^{k} = \frac{1}{\Lambda^k} \sum_{l=1}^{\Lambda^k} b_{\Lambda^k[(i-1)/\Lambda^k]+l}^{k-1},
\label{eq:b_i^k}
\end{eqnarray}
where ${\bf b}^0$ is initial bit-string array containing $0$ and $1$ elements, square brackets denote taking integer part. According to this expression at each iteration the whole array is divided into blocks of $\Lambda^k$ size, and elements within a block are substituted with the same value resulting from averaging all elements of the block (Fig.\ref{renorm}). Index $l$ denotes elements belonging to the same block. In our recent work~\cite{Dissimilarity} we have shown that in some cases it is enough to measure only part of the system to extract information about phase transitions. Following this way, for simplicity, we measured only 16 qubits in 19 site system to make the bit-string length an integer power of filter size $\Lambda$: $\log_{\Lambda} N \in {\mathbb N}$.

Dissimilarity between scales $k$ and $k+1$ is defined as
\begin{eqnarray}
\label{eq:Dissimilarity}
 {\cal D}_k =  |O_{k+1,k} - \frac12\left(O_{k, k}+O_{k+1, k+1}\right)|,
\end{eqnarray}
where $O_{m,n}$ is the overlap between vectors at scales $m$ and $n$:
\begin{eqnarray}
O_{m, n} = \frac{1}{L}\left({\bf b}^{m} \cdot {\bf b}^{n}\right).
\label{eq:overlap}
\end{eqnarray}
This expression can be considered as a modification of the structural complexity \cite{PNASComplexity} discussed above in the context of classical magnetic patterns to our quantum problem. 

There are two quantities of our principal interest: ${\cal D}_k$ that contains scale-resolved information on the pattern structure of the generated bit-string array and overall dissimilarity, ${\cal D} = \sum\limits_{k} {\cal D}_k$, where the sum goes over all the renormalization steps. $\cal D$ and $\left\{{\cal D}_k \right\}$ can be used for a unambiguous identification of a quantum state.

\begin{figure}[!t]
	\includegraphics[width=\columnwidth]{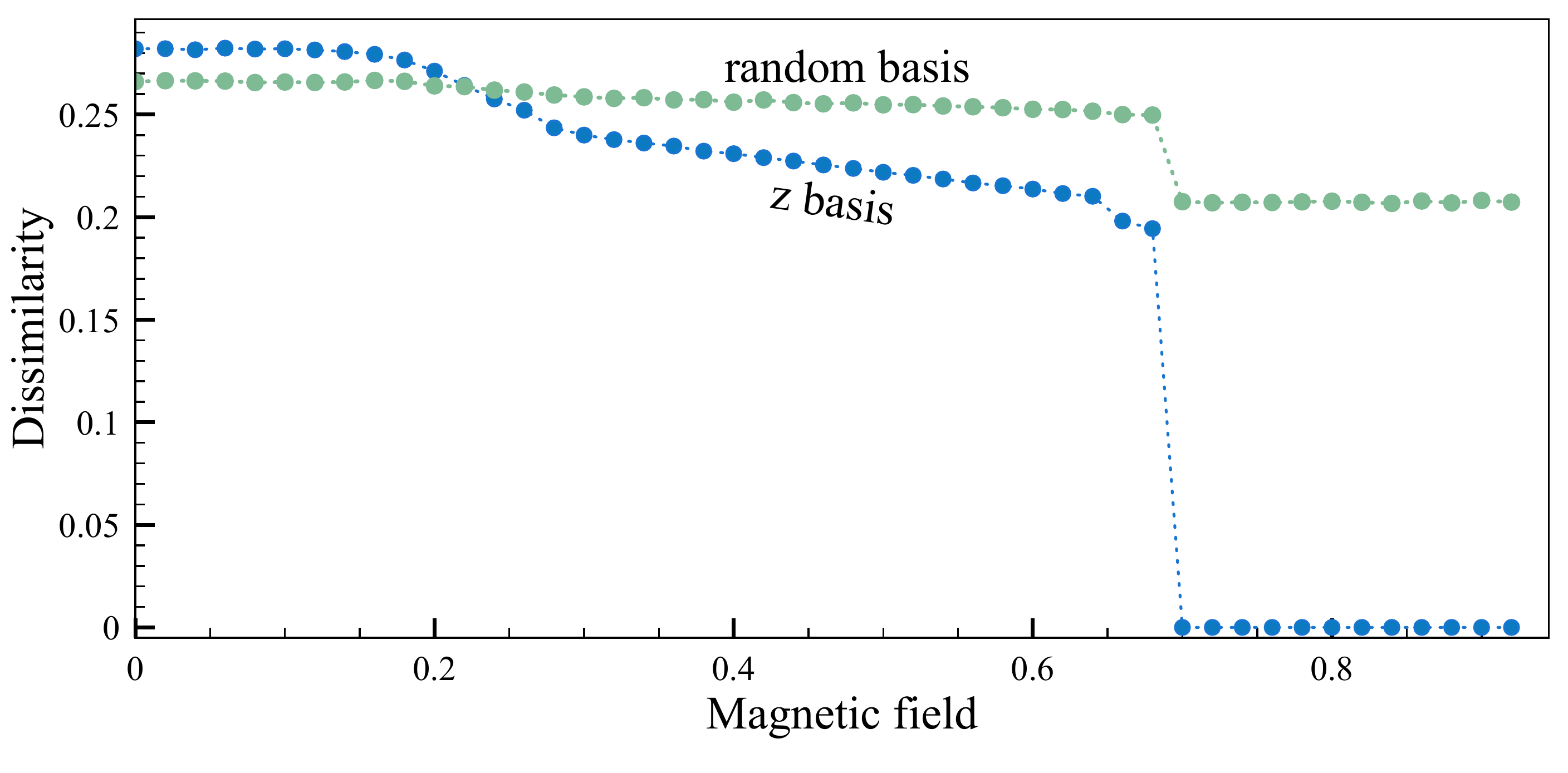}
	\caption{\label{diss_qsk} (Color online) The calculated dissimilarity as a function of the magnetic field.  }
\end{figure}

Fig.\ref{diss_qsk} gives the dissimilarity calculated in $\sigma^z$ and random bases for the ground state of the spin Hamiltonian, Eq.\ref{qHam} taken with $|{\bf D}_{ij}|$ = 1, $J_{ij}$ = 0.5 and $B$= 0.4.  In these calculations of ${\cal D}$ we used 16 of 19 bits from each measurement.  The calculated ${\cal D}^z$ reveals a smooth transition between spin spiral and skyrmion phases. On the other hand, the transition between skyrmion and ferromagnetic states is abrupt.  

\begin{figure}[!t]
	\includegraphics[width=\columnwidth]{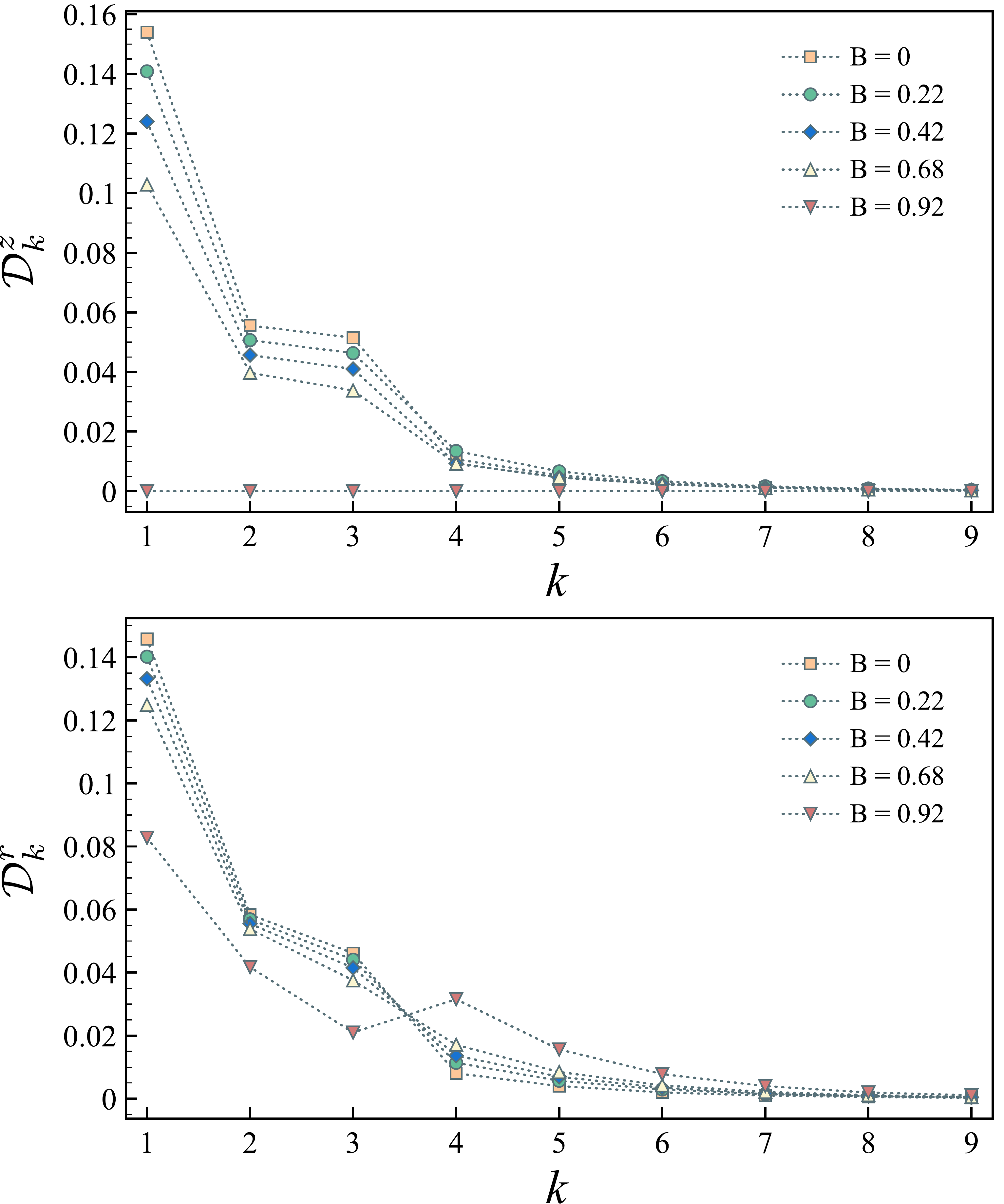}
	\caption{\label{diss_qsk_partial} (Color online) Partial dissimilarity of 19-site cluster calculated in (top) $\sigma^z$ and (bottom) random bases for different values of magnetic field $B$.}
\end{figure}

Figure~\ref{diss_qsk_partial} shows partial dissimilarities $\{\mathcal{D}_k\}$ calculated for 19-qubit system in $\sigma^z$ and random bases. Although partial dissimilarities of different non-collinear magnetic phases measured in $\sigma^z$ basis have similar shape, they still have different absolute values. Note, that in $\sigma^z$ basis the value of partial and total dissimilarity is equal to zero for ferromagnetic state. This happens because in $\sigma^z$ basis all measurements result in 0000\ldots{}0 or 1111\ldots{}1 depending on the direction of magnetic field which gives exactly zero for Eq.~\ref{eq:Dissimilarity}. The $\mathcal{D}^r$ demonstrates different behaviour for ferromagnetic phase, and, peak at $k = 4$ emerges when filter size $\Lambda^k$ becomes larger than the length of individual measurement bit-string effectively mixing different measurements.

Thus, the dissimilarity metric facilitates the search for phase boundaries of quantum models. Formally, this problem can be solved by choosing an appropriate order parameter, a correlation function characterized by a specific non-zero value for the definite range of model parameters \cite{qskyrm1, qskyrm2}. However, in this case one faces the problem that different phases cannot be described with a single order parameter. For instance, by using the scalar chirality we cannot distinguish between the cases of zero and high magnetic fields in the considered DMI magnet, Fig.\ref{qsk}. We would like to stress that one can consider the dissimilarity as a worthy alternative to other approaches for detecting phase boundaries in strongly correlated systems. First, the calculation of the dissimilarity is considerably cheaper in computational resources than traditional methods based on the calculation of the correlation functions. Then, the dissimilarity allows the accurate description of the phase boundaries in the cases when the choice of the order parameter is not obvious. It was demonstrated by the examples of the Shastry-Sutherland model of the orthogonal spin dimers and one-dimensional bond-alternating XXZ model hosting topological order \cite{Dissimilarity}.

\section{Conclusion}
To conclude, implementation of the machine learning methods considerably facilitates and accelerates theoretical characterization of non-collinear magnetic structures originated from the competition between anisotropic Dzyaloshinskii-Moriya and isotropic Heisenberg exchange interactions. As we have shown in this review by the example of the magnetic skyrmions the range of tasks to be solved is very wide, from constructing phase diagrams to estimating parameters of the parent Hamiltonians. In contrast to the traditional approach for description of magnetic systems that assumes accumulation of significant amount of the statistical data and calculations correlation functions of different orders, machine learning approaches do not require a complete information on the system in question. For instance, the temperature-magnetic-field phase diagrams can be constructed on the basis of few snapshots containing only $z$ projection of the spins. In this respect, we would like to emphasize a crucial role of the data preprocessing that can realize in different ways and forms. For instance, it could be simple sorting the magnetization vectors that considerably improves the quality of the supervised neural network classification. 

The procedure for estimating structural complexity of an object we described in the main text can be also considered as that performing a kind of preprocessing. It allows unsupervised identification of the phase boundaries of physical systems at a smooth varying external parameters such as temperature, magnetic field and others. As a result the complete phase space of the system in question can be divided onto separate phases, whose origin and properties could be further clarified with traditional and machine learning approaches. 

The same procedure for estimating pattern complexity can be applied in the case of quantum systems, for which one explores sequences of the bitstrings obtained from the projective measurements. As a prominent example we analyzed the pattern structure of the quantum ferromagnet with Dzyaloshinskii-Moriya interaction. Depending on the value of the external magnetic field one can observe different quantum phases including recently introduced quantum skyrmion one. Our results presented in this paper complete the picture of the quantum skyrmion properties and can be useful for characterization of such a state on the quantum computing devices.     

\section{Acknowledgements}
We thank Vladimir Dmitrienko for fruitful discussions.

\end{document}